\documentclass{pasj01}

\RequirePackage{mathpazo}
\RequirePackage[T1]{fontenc}

\newcounter{author}
\def\authorcount#1#2{\refstepcounter{author}\label{#1}
                     \altaffiltext{\ref{#1}}{#2}}

\def\Shibataprep{M. Shibata et al. in preparation}

\begin{document}
\SetRunningHead{T. Kato et al.}{Dwarf Nova Showing Both IW And and SU UMa-Type Features}

\Received{2021/04/02}%{yyyy/mm/dd}
\Accepted{202X/XX/XX}%{yyyy/mm/dd}

\title{BO Ceti: Dwarf Nova Showing Both IW And and SU UMa-Type Features}

\author{Taichi~\textsc{Kato},\altaffilmark{\ref{affil:Kyoto}*}
        Yusuke~\textsc{Tampo},\altaffilmark{\ref{affil:Kyoto}}
        Naoto~\textsc{Kojiguchi},\altaffilmark{\ref{affil:Kyoto}}
        Masaaki~\textsc{Shibata},\altaffilmark{\ref{affil:Kyoto}}
        Junpei~\textsc{Ito},\altaffilmark{\ref{affil:Kyoto}}
        Keisuke~\textsc{Isogai},\altaffilmark{\ref{affil:KyotoOkayama}}$^,$\altaffilmark{\ref{affil:Isogai2}}
        Hiroshi~\textsc{Itoh},\altaffilmark{\ref{affil:Ioh}}
        Franz-Josef~\textsc{Hambsch},\altaffilmark{\ref{affil:GEOS}}$^,$\altaffilmark{\ref{affil:BAV}}$^,$\altaffilmark{\ref{affil:Hambsch}}
        Berto~\textsc{Monard},\altaffilmark{\ref{affil:Monard}}$^,$\altaffilmark{\ref{affil:Monard2}}
        Seiichiro~\textsc{Kiyota},\altaffilmark{\ref{affil:Kis}}
        Tonny~\textsc{Vanmunster},\altaffilmark{\ref{affil:Vanmunster}}$^,$\altaffilmark{\ref{affil:Vanmunster2}}
        Aleksei~A.~\textsc{Sosnovskij},\altaffilmark{\ref{affil:CrAO}}
        Elena~P.~\textsc{Pavlenko},\altaffilmark{\ref{affil:CrAO}}
        Pavol~A.~\textsc{Dubovsky},\altaffilmark{\ref{affil:Dubovsky}}
        Igor~\textsc{Kudzej},\altaffilmark{\ref{affil:Dubovsky}}
        Tomas~\textsc{Medulka}\altaffilmark{\ref{affil:Dubovsky}}
}

\authorcount{affil:Kyoto}{
     Department of Astronomy, Kyoto University, Kyoto 606-8502, Japan}
\email{$^*$tkato@kusastro.kyoto-u.ac.jp}

\authorcount{affil:KyotoOkayama}{
     Okayama Observatory, Kyoto University, 3037-5 Honjo, Kamogatacho,
     Asakuchi, Okayama 719-0232, Japan}

\authorcount{affil:Isogai2}{
     Department of Multi-Disciplinary Sciences, Graduate School of
     Arts and Sciences, The University of Tokyo, 3-8-1 Komaba, 
     Meguro, Tokyo 153-8902, Japan}

\authorcount{affil:Ioh}{
     Variable Star Observers League in Japan (VSOLJ),
     1001-105 Nishiterakata, Hachioji, Tokyo 192-0153, Japan}

\authorcount{affil:GEOS}{
     Groupe Europ\'een d'Observations Stellaires (GEOS),
     23 Parc de Levesville, 28300 Bailleau l'Ev\^eque, France}

\authorcount{affil:BAV}{
     Bundesdeutsche Arbeitsgemeinschaft f\"ur Ver\"anderliche Sterne
     (BAV), Munsterdamm 90, 12169 Berlin, Germany}

\authorcount{affil:Hambsch}{
     Vereniging Voor Sterrenkunde (VVS), Oostmeers 122 C,
     8000 Brugge, Belgium}

\authorcount{affil:Monard}{
     Bronberg Observatory, Center for Backyard Astrophysics Pretoria,
     PO Box 11426, Tiegerpoort 0056, South Africa}

\authorcount{affil:Monard2}{
     Kleinkaroo Observatory, Center for Backyard Astrophysics Kleinkaroo,
     Sint Helena 1B, PO Box 281, Calitzdorp 6660, South Africa}

\authorcount{affil:Kis}{
     VSOLJ, 7-1 Kitahatsutomi, Kamagaya, Chiba 273-0126, Japan}

\authorcount{affil:Vanmunster}{
     Center for Backyard Astrophysics Belgium, Walhostraat 1A,
     B-3401 Landen, Belgium}

\authorcount{affil:Vanmunster2}{
     Center for Backyard Astrophysics Extremadura, e-EyE Astronomical Complex,
     06340 Fregenal de la Sierra, Badajoz, Spain}

\authorcount{affil:CrAO}{
     Federal State Budget Scientific Institution ``Crimean Astrophysical
     Observatory of RAS'', Nauchny, 298409, Republic of Crimea}

\authorcount{affil:Dubovsky}{
     Vihorlat Observatory, Mierova 4, 06601 Humenne, Slovakia}

%%% end:list of authors


\maketitle

\begin{abstract}
IW And stars are a recently recognized subgroup
of dwarf novae which are characterized by (often repetitive)
slowly rising standstills terminated by brightening,
but the exact mechanism for this variation is not yet
identified.
We have identified BO Cet, which had been considered
as a novalike cataclysmic variable, as a new member
of IW And stars based on the behavior in 2019--2020.
In addition to this, the object showed dwarf nova-type
outbursts in 2020--2021, and superhumps having a period 7.8\%
longer than the orbital one developed at least during
one long outburst.  This object has been confirmed
as an SU UMa-type dwarf nova with an exceptionally
long orbital period (0.1398~d).  BO Cet is thus
the first cataclysmic variable showing both SU UMa-type
and IW And-type features.
We obtained a mass ratio ($q$) of 0.31--0.34 from
the superhumps in the growing phase (stage A superhumps).
At this $q$, the radius of the 3:1 resonance,
responsible for tidal instability and superhumps,
and the tidal truncation radius are very similar.
We interpret that in some occasions this object
showed IW And-type variation when the disk size
was not large enough, but that the radius of
the 3:1 resonance could be reached as the result
of thermal instability.  We also discuss that
there are SU UMa-type dwarf novae above $q$=0.30, 
which is above the previously considered limit
($\sim$0.25) derived from numerical simulations and that 
this is possible since the radius of the 3:1 resonance 
is inside the tidal truncation radius.
We constrained the mass of the white dwarf larger
than 1.0$M_\odot$, which may be responsible for
the IW And-type behavior and the observed strength
of the He II emission.
The exact reason, however, why this object is unique in
that it shows both SU UMa-type and IW And-type features
is still unsolved.
\end{abstract}

\section{Introduction}

   Cataclysmic variables (CVs) are close binaries consisting
of a white dwarf (primary) and a mass-transferring
red or brown dwarf forming an accretion disk.
In some CVs, thermal instability in the accretion disk
causes outbursts and these CVs are called dwarf novae (DNe)
[see e.g. \citet{osa96review};
for general information of cataclysmic variables
and dwarf novae, see e.g. \citet{war95book}].
There are three classical subclasses of DNe:
SS Cyg stars with only normal outbursts, Z Cam stars
showing both normal outbursts and standstills and
SU UMa stars showing normal outbursts and superoutbursts,
during which which superhumps with
periods a few percent longer than
the orbital period ($P_{\rm orb}$) are present.

   Standstills in Z Cam stars are considered to be
equivalent to novalike (NL) stars, in which the disks are
thermally stable.  It is widely believed that
Z Cam stars have relatively high mass-transfer rates
and that the accretion disk is considered to be close to
the thermal stability.  It is considered
that a subtle variation in the mass-transfer rate
from the secondary causes transitions between
outbursting state and standstills \citep{mey83zcam}.
The exact mechanism, however, to produce standstills
is still not understood.

   In recent years, a group of Z Cam stars which show
unusual features has been identified \citet{sim11zcamcamp1}.
This group is currently called IW And
stars \citep{kat19iwandtype}.\footnote{
   They are also called ``anomalous Z Cam stars''
   (cf. \cite{ham14zcam}).
}
The features special to IW And stars are
(1) (quasi-)standstills are terminated by brightening,
not by fading as in ``textbook''
Z Cam stars, (2) the sequence is often very regular
with almost a constant recurrence time
(typically $\sim$30--100 d)
and (3) deep dips are sometimes seen following brightening.
Two numerical models have been proposed to explain
the IW And-type phenomenon: cyclic enhancement of
mass-transfer rates from the secondary \citep{ham14zcam}
and a modification of the standard thermal instability model
by considering of mass supply to the inner region
of the disk \citep{kim20iwandmodel}.
Although both models partially reproduced the IW And-type
light curves, they could not explain the radius
increase during (quasi-)standstills which are required
by observation (\cite{kim20kic9406652}; \Shibataprep).

   On the other hand, there is a rare DN, NY Ser,
which is at the same time an SU UMa star and a Z Cam star
\citep{kat19nyser}.  Two standstills in NY Ser were
terminated by superoutbursts, which indicates that
the disk radius can increase during standstills
and eventually reach the 3:1 resonance.  This object
provides the clearest evidence that the disk radius can
grow during standstills, which has not yet been
realized in numerical simulations.

   Here we report observations of BO Cet, which has shown
distinct states: long-lasting NL state, IW And-type
dwarf nova state and SU UMa-type superoutbursts.
There has been no other known object that shows
both IW And and SU UMa states.

\section{BO Cet}

   According to the General Catalog of Variable Stars
\citep{GCVSelectronic2011}, BO Cet was recognized as
an NL star by R. Remillard (1992), but without a solid
reference.  \citet{zwi95CVspec2} obtained a spectrum
and confirmed the NL nature.  \citet{rod07newswsex} performed
a more detailed spectroscopic study and classified
it as a non-eclipsing SW Sex star [see e.g. \citet{tho91bhlyn}
and \citet{rod07swsex} for the description of SW Sex stars].
An orbital period of 0.13980~d was obtained
by photometry by the members of the Center for Backyard
Astrophysics.\footnote{
   J. Patterson in 2002,
$<$http://cbastro.org/communications/news/messages/0274.html$>$.
} and was refined to be 0.13983~d by \citet{bru17CVphot1}.
\citet{lim21swsexmagnetic} obtained time-resolved
photometry and polarimetry, yielding a possible
spin period of 11.1~min by variation of
the circular polarization.  \citet{lim21swsexmagnetic}
detected a photometric period of 19.7~min.  Nearly
the same period was also detected radial velocities of
the H$\alpha$ line wings \citep{rod07newswsex}.

   During the early history of the research of BO Cet,
this object was considered as a NL star.  Indeed,
ASAS-3 \citep{ASAS3} and available AAVSO observations\footnote{
   $<$http://www.aavso.org/data-download$>$.
} did not show large variations in between 2001 and 2010.
During the course of a systematic study of cataclysmic
variables using Gaia DR2 \citep{GaiaDR2} and
All-Sky Automated Survey for Supernovae (ASAS-SN)
Sky Patrol (\cite{ASASSN}; \cite{koc17ASASSNLC}),
one of the authors (T.K.) noticed that this object
showed dwarf nova states in late 2013 and
2017 November to December (vsnet-chat 8150\footnote{
  $<$http://ooruri.kusastro.kyoto-u.ac.jp/mailarchive/vsnet-chat/8150$>$.
}, E-figure 1).
In 2020 June, the object was found to be in IW And
state since 2019 July (see subsection \ref{sec:bocet2019})
using the ASAS-SN data
(vsnet-alert 24387),\footnote{
  $<$http://ooruri.kusastro.kyoto-u.ac.jp/mailarchive/vsnet-alert/24387$>$.
} and a campaign was launched.

\section{Observations and Analysis}

   The CCD time-series observations were obtained in eight
locations between 2020 June and 2021 March
under a campaign led by the VSNET Collaboration \citep{VSNET}
comprised of 30-cm class telescopes.
The data analysis was performed just in the same way described
in \citet{Pdot} and \citet{Pdot6} and we mainly used
R software\footnote{
   The R Foundation for Statistical Computing:\\
   $<$http://cran.r-project.org/$>$.
} for data analysis.
In period analysis, we de-trending the data by using
both linear fitting and locally-weighted polynomial regression 
(LOWESS: \cite{LOWESS}) on 2--38 d segments 
to remove outburst trends depending on
the complexity of the light curve.
The times of superhumps maxima were determined by
the template fitting method as described in \citet{Pdot}.
The times of all observations are expressed in 
barycentric Julian days (BJD).

   We used phase dispersion minimization (PDM; \cite{PDM})
for period analysis and 1$\sigma$ errors for the PDM analysis
were estimated by the methods of \citet{fer89error}
and \citet{Pdot2}.

   We also obtained a snapshot spectrum between
BJD 2459157.9699 and 2459157.9856 using 
the Kyoto Okayama Optical Low-dispersion Spectrograph
with an Integral Field Unit (KOOLS-IFU; \cite{KOOLSIFU})
mounted on the 3.8-m telescope Seimei at Okayama Observatory,
Kyoto University \citep{Seimei}.
The wavelength coverage of VPH-Blue of KOOLS-IFU is 4200--8000\AA,
and its wavelength resolution is $R = \lambda/\Delta\lambda =$ 400-600.
The data reduction was performed using IRAF in
the standard manner (bias subtraction, flat fielding,
aperture determination, scattered light subtraction, 
spectral extraction, wavelength calibration with arc lamps and
normalization by the continuum).

\section{Results}

\subsection{2019-2020 season: IW And phase}\label{sec:bocet2019}

\begin{figure*}
  \begin{center}
    \FigureFile(150mm,70mm){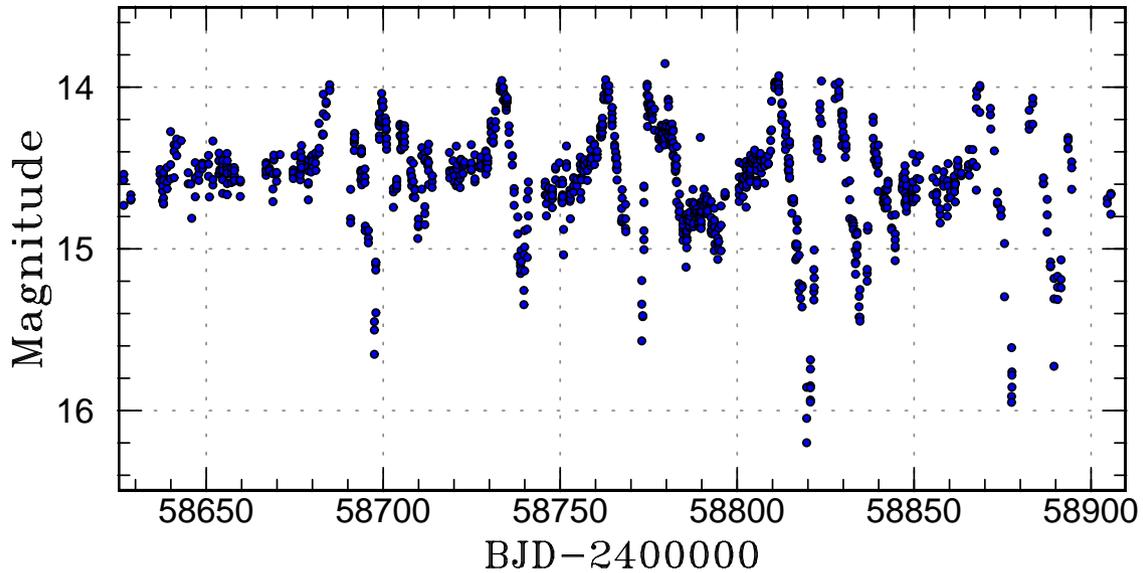}
  \end{center}
  \caption{The $g$-band light curve of BO Cet during 
  the 2019-2020 season from the ASAS-SN archive.
  This season started with a standstill or
  a NL state which likely persisted since 2016 December
  (observations were interrupted by solar conjunctions).
  until BJD 2458680.  This standstill or NL state was
  terminated by brightening in BJD 2458680--2458685.
  Two well-defined IW And-type cycles are seen
  between BJD 2458710 and 2458770: slowly rising
  standstills terminated by brightening followed by
  deep dips.  Other instances of termination of
  standstill by brightening are seen around BJD 2458810
  and 2458868.
  }
  \label{fig:bocet2019}
\end{figure*}

   The ASAS-SN light curve of the 2019-2020 season
(2019 May to 2020 February) is shown in
figure \ref{fig:bocet2019}.  
This season started with a standstill or
a NL state which likely persisted since 2016 December
(observations were interrupted by solar conjunctions)
until BJD 2458680.  This standstill or NL state was
terminated by brightening in BJD 2458680--2458685.
Two well-defined IW And-type cycles are seen
between BJD 2458710 and 2458770: slowly rising
standstills terminated by brightening followed by
deep dips.  Other instances of termination of
standstill by brightening are seen around BJD 2458810
and 2458868.  All these features support the classification
of BO Cet as an IW And star \citep{kat19iwandtype}.

\subsection{2020-2021 season: SU UMa/IW And phase}\label{sec:bocet2020}

\begin{figure*}
  \begin{center}
    \FigureFile(150mm,70mm){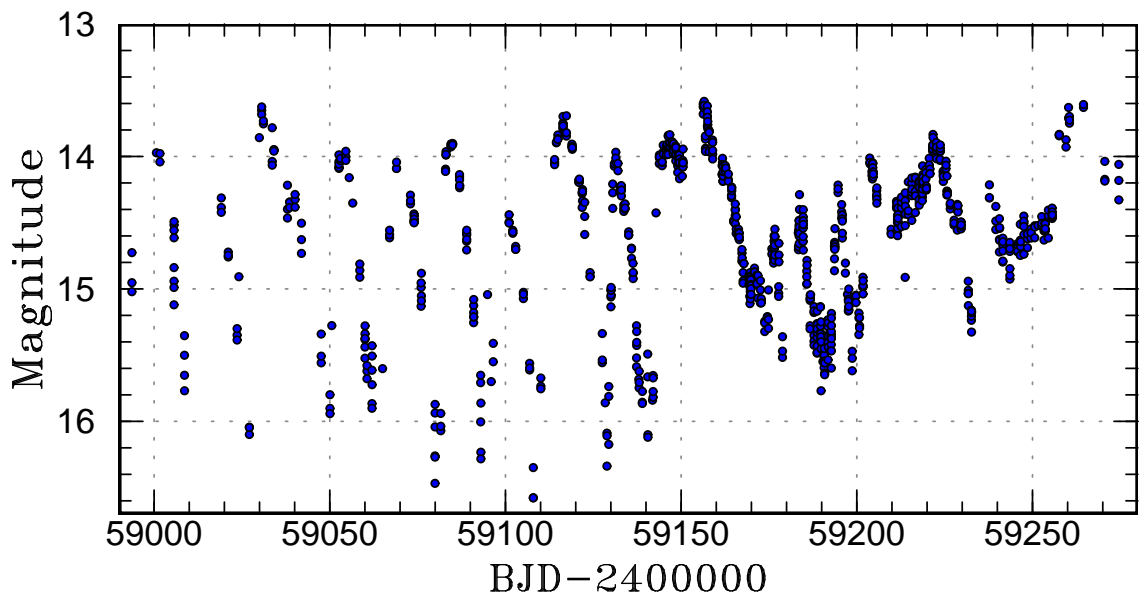}
  \end{center}
  \caption{The $g$-band light curve of BO Cet during 
  the 2020-2021 season from the ASAS-SN archive.
  The object showed dwarf nova-type outbursts
  until BJD 2459140 and no IW And-type feature was
  present.
  }
  \label{fig:bocet2020}
\end{figure*}

   The ASAS-SN light curve of the 2020-2021 season
(2020 May to 2021 February) is shown in
figure \ref{fig:bocet2020}.  The object showed
dwarf nova-type outbursts until BJD 2459140 and no
IW And-type feature was present.  Starting from
BJD 2459144, the object reached a plateau, which
was much brighter than standstills in the 2019-2020
season.  Time-resolved photometry recorded growing
superhumps during this plateau (subsection \ref{sec:SH}).
After then, the object further brightened.
After reaching the peak at $g$=13.6, the object
started fading slowly.  As shown later, this part
was identified as a superoutburst.  Superhumps
disappeared after this superoutburst and four small
outbursts occurred (BJD 2459176--2459205), followed
by a slowly rising standstill
(BJD 2459209--2459222).  Although less marked than in
the 2019-2020 season, this standstill was terminated
by weak brightening.  The object again entered
a slowly rising standstill (BJD 2459243--2459256)
terminated by brightening (BJD 2459257--).
Although the data were limited, this brightening
was probably accompanied by superhumps and was
likely a superoutburst.

   In conclusion, the 2020-2021 season was a combination
of SU UMa and IW And states.  Such behavior was first
recorded among all known CVs.

\subsection{Orbital signal}\label{sec:orb}

\begin{figure}
  \begin{center}
    \FigureFile(85mm,110mm){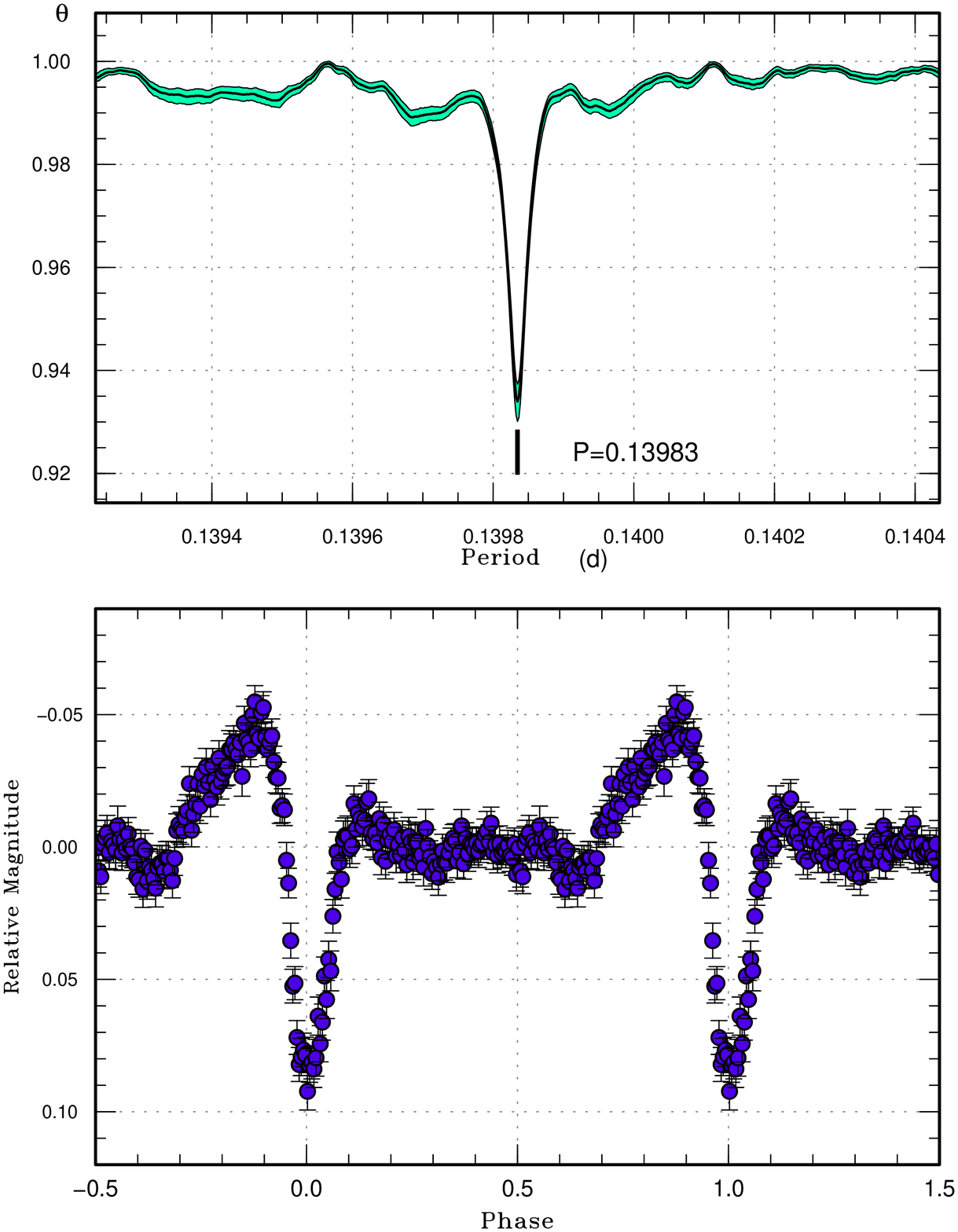}
  \end{center}
  \caption{Orbital signal of BO Cet.
     (Upper): PDM analysis.  The bootstrap result using
     randomly contain 50\% of observations is shown as
     a form of 90\% confidence intervals in the resultant 
     $\theta$ statistics.
     (Lower): Phase-averaged profile.  A shallow eclipse and
     an orbital hump are clearly seen. 1$\sigma$
     error bars are shown.}
  \label{fig:bocetorb}
\end{figure}

   Using all the data by VSNET Collaboration, we obtained
a photometric orbital period of 0.139835~d, 
which is in very good agreement with \citet{bru17CVphot1}.
The phase-averaged light curve clearly indicates
that the object is eclipsing (depth 0.09 mag)
accompanied by a prominent orbital hump
(figure \ref{fig:bocetorb}).
We have combined these data with ASAS-SN observations
during the NL (or long standstill) phase in 2014--2019
and improved the period by a Markov-Chain Monte Carlo (MCMC)
modeling of eclipses described in \citet {Pdot4}.
The eclipse ephemeris is
\begin{equation}
{\rm Min(BJD)} = 2459104.97794(9) + 0.13983546(3) E .
\label{equ:bocetecl}
\end{equation}

   The presence of a prominent orbital hump appears to be
inconsistent with the SW Sex-type classification
\citep{rod07newswsex}, since prominent orbital humps
are not usually seen in SW Sex stars \citep{tho91bhlyn}.

   We could not detect any coherent signals around
11.1~min or 19.7~min claimed by \citet{lim21swsexmagnetic}.
If one of these periods reflects the spin period of
the magnetized white dwarf (intermediate polar, IP),
such a period should have been easily detected as
a coherent signal in our observational condition
with dense and long-term coverage [see e.g.
\citet{kat15ccscl} for an example of detection of
an IP signal during a superoutburst under the similar
observational set-up].  We consider that the periods
detected by \citet{lim21swsexmagnetic} were more likely
quasi-periodic oscillations.

\subsection{Superhumps and mass ratio}\label{sec:SH}

\begin{figure}
  \begin{center}
    \FigureFile(85mm,70mm){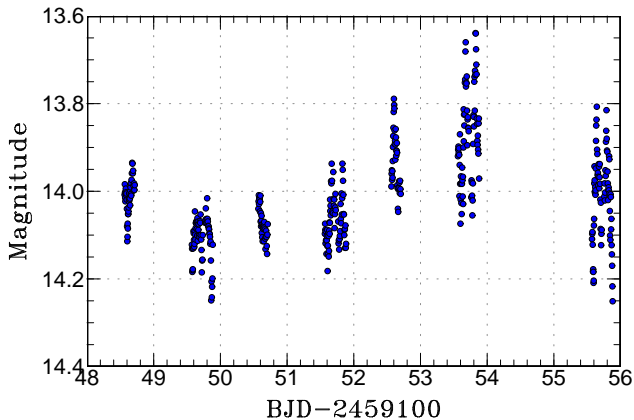}
  \end{center}
  \caption{Growing superhumps in BO Cet.
     The data were binned to 0.003~d.}
  \label{fig:bocetshlc}
\end{figure}

\begin{figure}
  \begin{center}
    \FigureFile(85mm,110mm){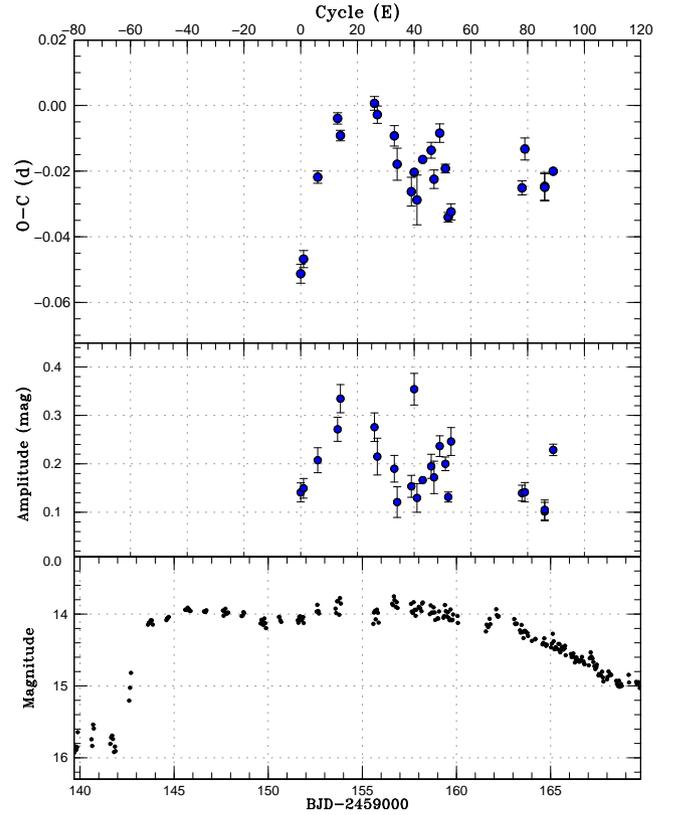}
  \end{center}
  \caption{$O-C$ diagram of superhumps in BO Cet.
     (Upper:) $O-C$ diagram.
     We used a period of 0.1507~d for calculating the $O-C$ residuals.
     (Middle:) Amplitudes of superhumps.
     (Lower:) Light curve.  The data were binned to 0.05~d.
     During the segment $0\le E \le 14$, the period was
     long and the amplitudes were rapidly growing.
     This segment corresponds to stage A.}
  \label{fig:bocetoc}
\end{figure}

   Superhumps appeared somewhere between BJD 2459144
and 2459151, and became prominent around BJD 2459154
(figure \ref{fig:bocetshlc}).
The times of superhump maxima are listed in E-table 1.  
Since the superhump period is long, it was rather
difficult to cover the sufficient phases of superhumps.
Although variations were apparent to the eyes,
some superhump maxima, particularly during the growing
phase, were not determined.  The $O-C$ curve and
amplitudes of the initial part, however, are sufficient
to show the growing phase (stage A, \cite{Pdot})
of superhumps (figure \ref{fig:bocetoc}).

\begin{figure*}
  \begin{center}
    \FigureFile(160mm,100mm){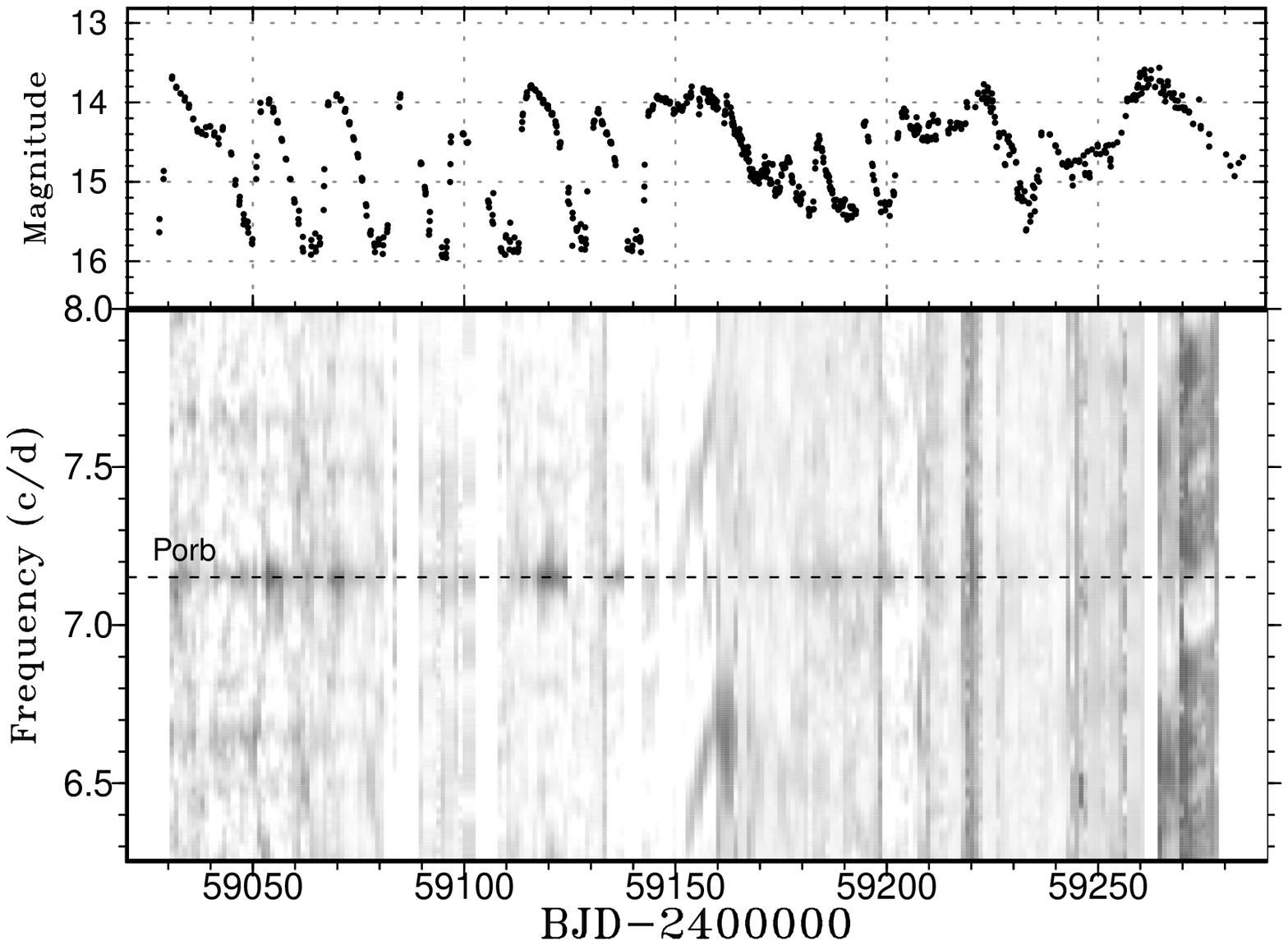}
  \end{center}
  \caption{Two-dimensional PDM analysis. From the top to the bottom, 
  (upper:) light curve; the data were binned to 0.1~d,
  (lower:) Two-dimensional PDM analysis.  4-d segments shifted
  by 1 d were analyzed.  Dark colors represent signals
  (lower $\theta$ in the PDM statistics).
  The orbital signal was present most of the time.
  The frequency around 6.7~c/d around BJD 2459160 represents
  superhumps.  The signal with rapidly increasing frequencies
  prior to this corresponds to stage A superhumps.
  The weak signal around 7.5--7.7~c/d in BJD 2459150--2459160
  is a one-day alias of the true frequency.}
  \label{fig:bocetspec2d}
\end{figure*}

   A two-dimensional PDM analysis is shown in
figure \ref{fig:bocetspec2d}.  This figure corresponds
to two-dimensional discrete Fourier transform or
least absolute shrinkage and selection operator (Lasso)
analysis employed in \citet{osa13v344lyrv1504cyg}.
In the case of BO Cet, PDM analysis gave a better
result due to the non-sinusoidal nature of superhumps
and nightly observational gaps.  The resolution of
PDM analysis is intermediate between Fourier transform
and Lasso analysis, and this result can be treated as
a slightly degraded version of Lasso analysis presented
in \citet{osa13v344lyrv1504cyg}.
The orbital signal was present most of the time.
The frequency around 6.7~c/d around BJD 2459160 represents
superhumps.  The signal with rapidly increasing frequencies
prior to this corresponds to stage A superhumps.
No signal of negative superhumps was present.

   A PDM analysis of the interval BJD 2459151--2459166
yielded a mean superhump period of 0.15069(3)~d, 7.8\%
longer than the orbital one (E-figure 2).

   We measured the period of superhumps during stage A
by using two different segments: 0.1529(3)~d for
BJD 2459152--2459154 and 0.1541(2)~d for BJD 2459150--2459154
with the PDM method.  These values correspond to
$\epsilon^* \equiv 1-P_{\rm orb}/P_{\rm SH}$ =
0.085(2) and 0.092(1), where
$P_{\rm SH}$ denotes the superhump period.
According to \citet{kat13qfromstageA}
the precession rate $\epsilon^*$ during stage A can be
directly translated into $q$ since the pressure effect
is considered to be neglected.  The above values
correspond to $q$=0.31(1) and 0.0335(4), respectively.
Considering the uncertainty to observationally determine
when superhumps started to grow, we adopted $q$=0.31--0.34
from this superhump analysis.

\subsection{Mass of the white dwarf}\label{sec:WDmass}

   We also modeled the eclipse profile assuming
an standard disk with a surface brightness proportional
to $r^{-0.75}$.  The disk radius was assumed to be
between an extreme values 0.30$A$, where $A$ is the binary
separation, and the tidal truncation radius.  The real
disk should be somewhere between them.
For $q$=0.31--0.34, we obtained
an inclination of $i$=71--73$^{\circ}$
(table \ref{tab:incl}).  This value is
not very dependent on the assumed power index
and we adopted a value of $i$=72(1)$^{\circ}$.
This value is significantly higher than
$i$=35--52$^{\circ}$ in \citet{rod07newswsex}.

\begin{table}[]
\caption{Orbital inclination determined by eclipse modeling}\label{tab:incl}
\begin{center}
\begin{tabular}{llll}
\hline
            &         & \multicolumn{2}{c}{mass ratio $q$} \\
            &         & 0.31 & 0.34  \\ \hline
Disk Radius & 0.30$A$ & 73$^{\circ}$ & 72$^{\circ}$ \\
            & 0.46$A$ & 72$^{\circ}$ & 71$^{\circ}$ \\
\hline
\end{tabular}
\end{center}
\end{table}

   The mass function is defined as
\begin{equation}
\frac{(M_1 \sin i)^3}{(M_1+M_2)^2} = \frac{P_{\rm orb} K_2^3}{2 \pi G},
\label{equ:massfunction}
\end{equation}
where $K_2$ is the semi-amplitude of the radial-velocity
variations of the secondary component.
In BO Cet, \citet{rod07newswsex} did not detect absorption
lines of the secondary and used an emission line
component corresponding to the motion of the secondary
in the Doppler tomogram.  This observed semi-amplitude of
the radial-velocity variations ($K_{\rm irr}$=262 km/s)
does not represent the motion of the center the mass of
the secondary, but that of the light center of
the irradiated region in the secondary's surface.
Thus $K_{\rm irr}$ is always smaller than $K_2$.

There is a relation
\begin{equation}
K_2 = \frac{K_{\rm irr}}{1-f(1+q)},
\label{equ:kirrtok2}
\end{equation}
where $f$ ($0 < f < 1$) represents the distance between
the center of the light center of the irradiated region
from the center the mass of the secondary in unit
of binary separation \citep{rod07newswsex}.
Assuming the theoretical limit
in which only the L$_1$ point is illuminated
(which condition is never actually achieved),
$f$ corresponds to the distance of the L$_1$ point
from the center the mass of the secondary ($R_{\rm L_2}$).
This extreme value is 0.383 for $q$=0.31 and
0.392 for $q$=0.34.  The lower limit for $f$ is achieved
when the hemisphere is maximally irradiated
and can be approximated by $f \simeq R_{\rm L_2}^2$
\citep{rod07newswsex}.
Using these values, we obtain $327 < K_2 < 542$ km/s.
The mass of the white dwarf is constrained as
$1.0 < M_1 < 4.7 M_\odot$.  Even taking the extreme limit
of the maximum irradiation, the white dwarf is more
massive than those in typical CVs, see
e.g. 0.82(3)$M_{\odot}$ in \citet{zor11SDSSCVWDmass}.
This result is consistent with a suggestion of
a massive white dwarf in the IW And-type postnova
BC Cas \citep{kat20bccas}.

\subsection{Spectroscopy}

\begin{figure*}
  \begin{center}
    \FigureFile(130mm,60mm){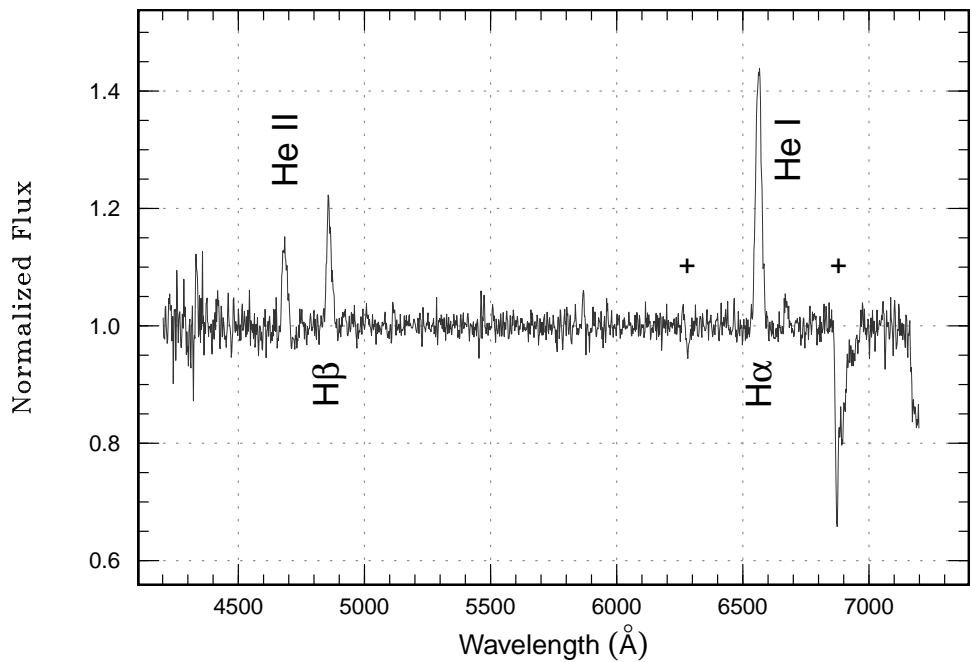}
  \end{center}
  \caption{Spectrum of BO Cet taken during the superoutburst
  on 2020 November 4.  The ``+'' signs represent telluric
  absorptions.
  }
  \label{fig:BOCet_all_normalized}
\end{figure*}

   The result of snapshot spectroscopy is shown in
figure \ref{fig:BOCet_all_normalized}.
the equivalent widths (EW) and full widths at half maximum
(FWHM) of strong lines are summarized in
table \ref{tab:spec}.

   In the spectrum of \citet{zwi95CVspec2} during
an apparent NL state, BO Cet showed a relatively strong
He II emission.  Compared to this, the current
observation seems to show a slightly enhanced
He II emission, but the difference is subtle.
A slight enhancement was probably due to
the enhanced mass-accretion rate to the white dwarf
during the superoutburst.
The strong He II emission may be the result of
a white dwarf in BO Cet more massive than in
ordinary CVs (subsection \ref{sec:WDmass}) [see
\citet{kim18j1621} examples of He II emission in DNe
containing massive white dwarfs].

   \citet{rod07newswsex} indicated that the H$\alpha$
emission is singly peaked most of the time.
The current observation also confirmed the singly peaked
H$\alpha$ emission despite the relatively high
orbital inclination.  We know that not all high-inclination
dwarf novae show doubly peaked emission lines
during superoutbursts, the best example being
the WZ Sge-type dwarf nova V455 And
(\cite{nog09v455andspecproc}; Y. Tampo et al. in preparation).  
We consider that there may be a cause other than
the SW Sex-type phenomenon producing singly peaked
emission lines in the light of the lack of
signature of an SW Sex star in BO Cet in the orbital
light curve (subsection \ref{sec:orb}).

\begin{longtable}{cccccc}
  \caption{Equivalent Width (EW) and Full Width at Half Maximum (FWHM)
  of our spectrum}\label{tab:spec}
  \hline
    \multicolumn{2}{c}{H$\alpha$} & \multicolumn{2}{c}{H$\beta$} &
\multicolumn{2}{c}{He II 4686\AA}\\
     EW & FWHM & EW & FWHM & EW & FWHM \\
     & (\AA) &  & (\AA) &  & (\AA) \\
    \hline
    \hline
\endfirsthead
    12.4(1) & 26.3(2) & 4.7(1) & 22.0(5) & 3.2(1) & 21.4(5) \\
\hline
\end{longtable}

\section{Discussion}

\subsection{Mass ratio and SU UMa-type phenomenon}\label{sec:massratio}

   Although the $q$ value (0.31--0.34) is expected
normal for an binary with $P_{\rm orb}$=0.13983546 d,
this mass ratio is surprisingly large for an SU UMa
star, since the 3:1 resonance is considered to be
achieved only for $q<0.25$ \citep{whi88tidal}
or in extreme cases of 0.33 under temporary reduction
of the mass transfer from
the secondary \citep{mur00SHintermediateq}.
The recent results by \citet{wak21asassn18aan} also
support that this limit is difficult to break,
and that tidal truncation provides a hard limit
for a disk to grow.  However, since the secondary of
BO Cet is not reported to be anomalous as in ASASSN-18aan
(\cite{wak21asassn18aan}; the lack of the secondary
eclipse in BO Cet also excludes a luminous secondary
as in ASASSN-18aan), the present results
indicate that the limits presented by the past
smoothed-particle hydrodynamics (SPH) simulations
may not be as stringent as previously considered.

   In fact, there has been yet another case of
a super-long-period SU UMa star
(MisV1448, $P_{\rm SH}$=0.237~d, vsnet-alert 24912)
most likely having $q > 0.30$.
Combined with ASASSN-14ho \citep{kat20asassn14ho},
which had outburst characteristics very similar to
MisV1448 (but yet without superhump detection), SU UMa stars
above $q$=0.30 had simply been missed by observations and
may have simply been underestimated: observers
tend to observe high-amplitude dwarf novae to establish
the SU UMa-type nature while systems with high $q$ have
smaller outburst amplitudes; Z Cam stars usually have not been
selected for time-resolved photometry in search of
superhumps since it has been generally accepted that
Z Cam-type and SU UMa-type categories are exclusive.

   The treatment in the past SPH simulations also may have
been responsible for this apparent discrepancy.
According to \citet{oya21MHDecc}, the viscosity in the disk, 
which is a reflection of the poloidal magnetic flux, 
determines the growth of eccentricity.  It was possible
that the past SPH simulations did not consider such
an effect properly.  Even in $q$=0.31--0.34, the 3:1 resonance
is within the tidal truncation radius (see subsection
\ref{sec:superoutbursts}) and it is not logically impossible
to excite the 3:1 resonance under these high $q$ values.

\subsection{Origin of superoutbursts}\label{sec:superoutbursts}

   We compared the radius of the 3:1 resonance
($r_{\rm 3:1}=3^{-2/3} (1+q)^{-1/3} A$)
and tidal truncation radius.
We used the approximated formula of tidal truncation radius
$r_{\rm tidal}=0.60/(1+q) A$ for 0.03$<q<$1
given by \citet{pac77ADmodel}.

   For $q$=0.31, these values are $r_{\rm 3:1}=0.440A$
and $r_{\rm tidal}=0.46 A$.  For $q$=0.34, they are
0.436$A$ and 0.45$A$, respectively.
It is worth noting that $r_{\rm 3:1}$ and $r_{\rm tidal}$
are critically close in the parameter of BO Cet.
The disk radius during standstills in IW And stars
is known to increase at least in two studies
(\cite{kim20kic9406652}; \Shibataprep).
The same is true for the only previously known
Z Cam/SU UMa star NY Ser \citep{kat19nyser}.
Furthermore, direct observational determination of
the disk radius by \Shibataprep \, indicates that IW And-type
standstills are terminated by brightening when
the disk radius eventually reaches
the tidal truncation radius, just like how
the radius 3:1 resonance plays in SU UMa stars.
It looks likely that there exists
a mechanism of effective removal of angular momentum
at the tidal truncation radius [see also a discussion
in subsection 4.1 in \citet{kim20kic9406652}].

   In BO Cet, this removal of angular momentum
probably starts to work when the disk expands
to the tidal truncation radius during a standstill,
and a standstill ends up with IW And-type
brightening often followed by a deep dip.

   In figure \ref{fig:bocet2019}, IW And-type
brightening at the end of standstills reached
a constant magnitude of 14.0.\footnote{
  The ASAS-SN and our CCD magnitudes can be regarded as
  the same for the disk-dominated ($B-V \sim 0$) system
  BO Cet.
}  This fixed brightness can be regarded
as a signature of the disk reaching the tidal truncation radius
and the associated effective removal of the angular momentum
working.  In the DN phase in 2020 before
BJD 2459140, however, this limit of 14.0 mag was
surpassed several times (figure \ref{fig:bocet2020}),
suggesting that the disk expanded beyond
the tidal truncation radius.  This is likely due to
thermal instability causing the disk to expand
due to increased viscosity under the condition of
the angular momentum conservation.  In the IW And phase,
such a large variation of viscosity is not expected
during standstills and the absence of superoutbursts
starting from standstills appears to be explained
by the inability of the disk expanding beyond
the tidal truncation radius.
Even if the disk radii apparently exceeded
the tidal truncation radius (and probably reaching
the radius of the 3:1 resonance) during some outbursts
in the DN phase, only one or two outbursts
eventually became superoutbursts.  This was probably
because tidal instability at the 3:1 resonance requires
time to develop (cf. \cite{lub91SHa}; \cite{lub91SHb})
and only long-lasting outbursts could eventually
develop tidal instability to manifest the SU UMa-type
feature.

\subsection{Unsolved questions}\label{sec:unsolved}

   Although the phenomenon observed in BO Cet appears
to be explained by a chance (nearly) superposition 
of $r_{\rm 3:1}$ and $r_{\rm tidal}$, there remains
a number of unknowns.  We summarize them as hints
to future studies in understanding of the accretion disks
in dwarf novae.

\begin{itemize}

\item \textbf{Relation between IW And and SU UMa phenomena.}

Both IW And and SU UMa phenomena are powered
by the presence of a characteristic disk radius
at which either IW And-type brightening or
SU UMa-type superoutburst occurs.
Do the underlying mechanisms in these systems have
something in common?

\item \textbf{Interaction of the phenomena at $r_{\rm 3:1}$
and $r_{\rm tidal}$.}

In SU UMa stars with large $q$, the development
of superhumps are known to be slow (\cite{Pdot6};
\cite{kat16v1006cyg}).  Could this be a result
of competition or interference between tidal instability
at the radius of the 3:1 resonance and tidal truncation?
Alternately, could the 3:1 resonance affect
the behavior when the disk radius reaches
the tidal truncation radius in systems having
similar $q$ to BO Cet?

\item \textbf{Why not many systems with intermediate $q$
show superoutbursts?}

Not all systems with intermediate $q$ (0.25--0.35)
have shown superoutbursts.
What is the difference between BO Cet and other systems?
Is the case of BO Cet related to
the condition allowing IW And-type phenomenon
to occur, which, however, is still poorly understood?

\end{itemize}

\section{Summary}

   We observed BO Cet, which had been considered
as a novalike cataclysmic variable.
During the 2020--2021 season, the object showed
dwarf nova-type outbursts and superhumps having
a period 7.8\% longer than the orbital one developed
at least during one long outburst.
This object has thus been confirmed
as an SU UMa-type dwarf nova with an exceptionally
long orbital period [0.13983546(3)].

   Furthermore, we found that the object showed 
IW And-type variation in 2019--2020 in the archival
data, which is characterized by (often repetitive) slowly
rising standstills terminated by brightening.
This object is thus the first cataclysmic variable
showing both SU UMa-type and IW And-type features.

   We found that the object is eclipsing and estimated
the orbital inclination to be 72(1)$^{\circ}$.
Using the published radial-velocity study,
we constrained the mass of the white dwarf
larger than 1.0$M_\odot$.  This supports the previous
claim that the IW And-type phenomenon (in some objects)
may be associated with the high mass of the white dwarf.
The massive white dwarf may also be responsible
for the strong He II line in the spectra.

   We obtained a mass ratio of 0.31--0.34 from
the superhumps in the growing phase (stage A superhumps).
At this mass ratio, the radius of the 3:1 resonance,
responsible for tidal instability and superhumps,
and the tidal truncation radius are very similar
and might lead to the unusual behavior in this object.

   We interpret that in some occasions this object
showed IW And-type variation when the disk size
was not large enough but that the radius of
the 3:1 resonance could be reached as the result
of thermal instability.  The reason, however, why
this object is unique in that it shows both
SU UMa-type and IW And-type features is still
unsolved.

\section*{Acknowledgments}

The author is particularly grateful to the ASAS-3 and
ASAS-SN team for making their data available to the public.
We acknowledge with thanks the variable star
observations from the AAVSO International Database contributed
by observers worldwide and used in this research.
This research has made use of the International Variable Star Index 
(VSX) database, operated at AAVSO, Cambridge, Massachusetts, USA.
We also benefited from the data by Filipp Romanov.
This work was supported by JSPS KAKENHI Grant Number 21K03616.

\end{document}